\documentclass{elsart}

\usepackage{natbib}
\usepackage{graphicx}
\usepackage{amssymb}

\begin{document}
\begin{frontmatter}

\title {Production and Evolution of Millisecond X-ray and Radio Pulsars\thanksref{ack}}

\thanks[ack]{This research was supported in part by NASA grant NAG~5-12030, NSF grant AST~0098399, and the funds of the Fortner Endowed Chair at the University of Illinois.}

\author {Frederick K. Lamb}

\address {Center for Theoretical Astrophysics, Department of
  Physics, and Department of Astronomy, University of Illinois at
  Urbana-Champaign, 1110 W. Green Street, Urbana, IL 61801-3080,
  USA; fkl@uiuc.edu}

\begin {abstract}
Observations using the \textit {Rossi X-ray Timing Explorer} (\textit {RXTE}) have discovered dozens of accreting neutron stars with millisecond spin periods in low-mass binary star systems. Eighteen are millisecond X-ray pulsars powered by accretion or nuclear burning or both. These stars have magnetic fields strong enough for them to become millisecond rotation-powered (radio) pulsars when accretion ceases. Few, if any, accretion- or rotation-powered pulsars have spin rates higher than 750~Hz. There is strong evidence that the spin-up of some accreting neutron stars is limited by magnetic spin equilibrium whereas the spin-up of others is halted when accretion ends. Further study will show whether the spin rates of some accretion- or rotation-powered pulsars are or were limited by emission of gravitational radiation.
\end {abstract}

\begin{keyword}
accretion, accretion disks \sep pulsars \sep X-rays: binaries \sep stars: neutron
\PACS 95.85.Nv \sep 97.10.Gz \sep 97.60.Gb \sep 97.60.Jd \sep 97.80.Jp
\end{keyword}
\end{frontmatter}

% main text
\section {Observed Properties of Millisecond X-ray and Radio Pulsars
\label {MSPs}}

In this brief review, a pulsar is considered a millisecond pulsar (MSP) if its spin period $P_{s}$ is $<10$~ms (spin frequency $\nu_{s}>100$~Hz). Seven accretion-powered and 13 nuclear-powered X-ray MSPs are now known (see Table~1). All are accreting neutron stars in low-mass X-ray binary systems (LMXBs). Two of the accretion-powered MSPs become nuclear-powered MSPs during thermonuclear X-ray bursts. Kilohertz quasi-periodic oscillations (QPOs) have so far been detected in the accretion-powered X-ray emission of more than two dozen neutron stars in LMXBs, with frequencies ranging from $\sim\,$100~Hz to $\sim\,$1300~Hz (Lamb 2003; van der Klis 2006). They have been seen in  8 of the 13 known nuclear-powered MSPs and 2 of the 7 known accretion-powered MSPs (see Table~1).

\begin {table}[t!]
\begin {center}
\begin {minipage}{140 mm}
\caption {Nuclear- and Accretion-Powered Millisecond X-ray Pulsars}
\end {minipage}
\begin {minipage}{140 mm}
\vspace {3pt}
\begin {tabular}{cll}
\hline
\hline
\noalign{\kern 2pt}
$\nu_{\rm spin}$~(Hz)$^a$  &Object   &References \\
\noalign{\kern 2pt}
\hline
\noalign{\kern 0pt}
619\ \ NK\quad\qquad      &\hbox{4U~1608$-$52}               &Hartman et al.\ 2003\\
\noalign{\kern -6pt}
601\ \ NK\quad\qquad      &\hbox{SAX~J1750.8$-$2900}         &Kaaret et al.\ 2002\\
\noalign{\kern -6pt}
598\ \ A\qquad\qquad      &\hbox{IGR~J00291$+$5934}          &Markwardt, Swank \& Strohmayer 2004\\
\noalign{\kern -6pt}
589\ \ N\qquad\qquad      &\hbox{X~1743$-$29}                &Strohmayer et al.\ 1997\\
\noalign{\kern -6pt}
581\ \ NK\quad\qquad      &\hbox{4U~1636$-$53}               &Zhang et al.\ 1996, Strohmayer et al.\ 1998\\
\noalign{\kern -6pt}
567\ \ N\qquad\qquad      &\hbox{X~1658$-$298}               &Wijnands, Strohmayer \& Franco 2001\\
\noalign{\kern -6pt}
549\ \ NK\quad\qquad      &\hbox{Aql~X-1}                    &Zhang et al.\ 1998\\
\noalign{\kern -6pt}
524\ \ NK\quad\qquad      &\hbox{KS~1731$-$260}              &Smith, Morgan \& Bradt 1997\\
\noalign{\kern 2pt}
\hline
\noalign{\kern 0pt}
435\ \ A\qquad\qquad      &\hbox{XTE~J1751$-$305}            &Markwardt et al.\ 2002\\
\noalign{\kern -6pt}
410\ \ N\qquad\qquad      &\hbox{SAX~J1748.9$-$2021}         &Kaaret et al.\ 2003\\
\noalign{\kern -6pt}
401\ \ ANK\ \             &\hbox{SAX J1808.4$-$3658\quad}    &Chakrabarty \& Morgan 1998\\
\noalign{\kern -6pt}
                          &                                  &Wijnands \& van der Klis 1998\\
\noalign{\kern -6pt}
                          &                                  &Chakrabarty et al.\ 2003\\
\noalign{\kern -6pt}
                          &                                  &Wijnands et al.\ 2003\\
\noalign{\kern -6pt}
\ 377\ \ A\qquad\qquad    &\hbox{HETE~J1900.1$Ð$2455}        &Morgan, Kaaret \& Vanderspek 2005\\
\noalign{\kern -6pt}
363\ \ NK\quad\qquad      &\hbox{4U~1728$-$34}               &Strohmayer et al.\ 1996\\
\noalign{\kern -6pt}
330\ \ NK\quad\qquad      &\hbox{4U~1702$-$429}              &Markwardt, Strohmayer \& Swank 1999\\
\noalign{\kern -6pt}
314\ \ AN\quad\qquad      &\hbox{XTE~J1814$-$338}            &Strohmayer et al.\ 2003\\
\noalign{\kern -6pt}
\ 270\ \ N\qquad\qquad    &\hbox{4U~1916$-$05}               &Galloway et al.\ 2001\\
\noalign{\kern -6pt}
191\ \ AK\quad\qquad      &\hbox{XTE~J1807.4$-$294}          &Linares et al.\ 2005\\
\noalign{\kern -6pt}
\ 185\ \ A\qquad\qquad    &\hbox{XTE~J0929$-$314}\ \ \ \     &Galloway et al.\ 2002\\
\noalign{\kern -6pt}
\ \ 45\ \ N\qquad\qquad   &\hbox{EXO~0748$-$676}\ \ \ \      &Villarreal \& Strohmayer 2004\\
\noalign{\kern 4pt}
\hline
\noalign{\kern 4pt}
\end {tabular}
\end {minipage}
\begin {minipage}{140 mm}
{$^a$Spin frequency inferred from periodic or nearly periodic X-ray oscillations. A:~accretion-powered millisecond pulsar. N: nuclear-powered millisecond pulsar. K: kilohertz QPO source. See text for details.}
\end {minipage}
\end {center}
\end {table}

The discovery of a pair of kilohertz QPOs in the 401~Hz accretion- and nuclear-powered X-ray pulsar \mbox {SAX~J1808.4$-$3658} (Chakrabarty et al.\ 2003; Wijnands et al.\ 2003), of a pair of kilohertz QPOs in the 191~Hz accretion-powered X-ray pulsar \mbox {XTE~J1807.4$-$294} (Linares et al.\ 2005), and of burst oscillations in the 314~Hz accretion-powered X-ray pulsar \mbox {XTE~J1814$-$338} (Strohmayer et al.\ 2003) has confirmed the relationship between these three distinct types of high-frequency X-ray oscillations. 

Except during the first few seconds of some bursts, the nuclear-powered oscillations of {SAX~J1808} and \mbox {XTE~J1814} have the same frequency and phase as the accretion-powered oscillations. These observations establish beyond any doubt that (1)~the nuclear- and accretion-powered oscillations are both produced by spin modulation of the X-ray flux from the stellar surface and (2)~the magnetic fields of these stars are strong enough to channel the accretion flow and enforce corotation of gas heated by nuclear burning, which requires fields $\gtrsim10^{7}$~G (Miller, Lamb \& Psaltis 1998; Lamb \& Yu 2005). Conversely, the nearly sinusoidal waveforms and low amplitudes of these oscillations indicate that the surface magnetic fields of these neutron stars are less than $\sim10^{10}$~G (Psaltis \& Chakrabarty 1999). These results confirm that the frequencies of both the accretion- and the nuclear-powered X-ray oscillations are the spin frequencies of these neutron stars.

None of the 18 currently known accretion- and nuclear-powered X-ray pulsars have spin frequencies higher than 619~Hz (see Table~1). There is no known observational bias against detecting such pulsars. This indicates that few, if any, such pulsars have spin frequencies higher than 750~Hz (Chakrabarty et al.\ 2003; M.~C. Miller 2005, personal communication). Although too few such pulsars have been observed to establish their spin distribution accurately, the spins of the known MSPs nevertheless provide valuable information about their production and evolution, as discussed below.

A weak-field accreting neutron star typically produces a pair of strong kilohertz QPOs, with frequencies that vary by as much as a factor of five as the star's X-ray flux varies by a factor of two or three (see Miller et al.\ 1998; Lamb 2003; Lamb 2005; Lamb \& Miller 2006; van der Klis 2006). As the frequencies of the two kilohertz QPOs vary by hundreds of Hertz, their separation $\Delta\nu_{\rm QPO}$ remains roughly constant. All the kilohertz QPO pairs detected in X-ray pulsars have $\Delta\nu_{\rm QPO}$ approximately (sometimes very accurately) equal to either $\nu_{s}$ or $0.5\,\nu_{s}$. For example, the kilohertz QPO pair discovered in XTE~J1807.4 has $\Delta\nu_{\rm QPO} = \nu_{s}$ (Linares et al.\ 2005) whereas that discovered in SAX~J1808.4 has $\Delta\nu_{\rm QPO} = 0.5\,\nu_{s}$ (Wijnands et al.\ 2003). These findings show that the spin of the star plays a central role in generating its kilohertz QPO pair. This is possible only if the neutron stars that produce kilohertz QPOs have magnetic fields $\gtrsim 10^{8}$~G  (Miller et al.\ 1998; Lamb \& Miller 2001; Lamb \& Miller 2006). The central role played by the star's spin frequency means that information about its spin frequency can be extracted from the frequencies of its kilohertz QPOs, even if accretion- or nuclear-powered oscillations have not yet been detected.

Altogether, more than two dozen accreting neutron stars have spin rates and magnetic fields high enough for them to become rotation-powered radio-emitting MSPs when accretion ceases, supporting the hypothesis (Alpar et al.\ 1982; Radhakrishnan \& Srinivasan 1982) that such neutron stars are the progenitors of the radio MSPs. Many new rotation-powered radio MSPs have been discovered in recent years and dozens are now known (see Lorrimer 2005). Eighty binary and millisecond radio pulsars have been associated with the Galactic disk. More than one hundred radio pulsars have been discovered in 24 globular clusters in the Galaxy (Lorrimer 2005). In November 2005, a radio pulsar was discovered in Terzan~5 with a spin frequency of 716~Hz (Hessels et al.\ 2006), higher than the 642~Hz spin frequency of the first radio MSP discovered (Backer at al.\ 1982).

\section {Production of Millisecond X-ray and Radio Pulsars}

\textit {Production and spin evolution of accretion-powered millisecond X-ray pulsars}.---The timescale on which accretion from a disk doubles the spin rate of a slowly rotating magnetic neutron star is (Ghosh \& Lamb 1979b; Ghosh \& Lamb 1992)
\begin {equation}
t_{s} \equiv 2\pi\nu_{s}I/[{\dot M} (GMr_m)^{1/2}] \sim 10^8 \, {\rm yr}\, \left( \frac{\nu_{s}}{\rm 300~Hz} \right)  \left( \frac{\dot M} {0.01{\dot M}_E} \right)^{-1+\alpha/3}\;. \label{t-s}
\end {equation}
\vskip-10 pt
Here $\nu_{s}$, $M$, and $I$ are the star's spin frequency, mass, and moment of inertia, ${\dot M}$ is the mass accretion rate onto the star (not the mass transfer rate), ${\dot M}_E$ is the mass accretion rate that produces the Eddington luminosity, and $r_m$ is the angular momentum coupling radius. In the final expression on the right-hand side of equation~(\ref{t-s}), the weak dependence of $t_{s}$ on $M$, $I$, and the star's magnetic field has been neglected; $\alpha$ is 0.23 if the inner disk is radiation-pressure-dominated (RPD) or 0.38 if it is gas-pressure-dominated (GPD).

The first papers to argue that pulsing X-ray stars are accreting magnetic neutron stars also argued that accretion of angular momentum would spin such stars up until their angular frequencies are approximately equal to the orbital angular frequency at the disk-magnetosphere coupling radius (see Lamb, Pethick \& Pines 1973). Accretion from a disk drives the neutron star's spin rate upward or downward on the timescale~(\ref{t-s}) until it reaches the ``magnetic spin equilibrium'' period $P_{\rm eq}$. Subsequent two-dimensional semi-analytical MHD calculations (Ghosh \& Lamb 1979a; Ghosh \& Lamb 1979b; Ghosh \& Lamb 1992) supported the conclusion that the torque on a magnetic neutron star accreting from a disk vanishes at a certain period and provided a more accurate value of this period. Recent two- and three-dimensional numerical MHD calculations (Romanova et al.\ 2002) have provided further support. 

By 1982, spin-up by accretion in binary systems was being considered to explain the short-period ($\sim\,$60--100~ms) rotation-powered radio pulsars that had been discovered (see Taylor \& Stinebring 1986). That year Alpar et al.\ (1982) and Radhakrishnan \& Srinivasan (1982) argued that some neutron stars have high enough accretion rates and weak enough magnetic fields that they can be spun up to periods as short as a millisecond. That same year Backer et al.\ (1982) discovered a rotation-powered pulsar with a 1.56~ms period (\mbox{PSR~1937$+$21}).

Magnetic neutron stars formed in close binary systems are thought to be spun down initially, by emission of particles and electromagnetic radiation, and then spun up by accreting angular momentum when mass transfer occurs. This ``recycling'' process produces weak-field neutron stars with high spin rates. In order to obtain a definite equilibrium spin period (``spin-up line''), Alpar et al.\ and Radhakrishnan \& Srinivasan assumed that the accretion rate at the end of mass transfer is the Eddington rate, even for low-mass binary systems. This is the highest accretion rate that is possible, but it is not the accretion rate expected at the end of mass transfer in the relevant binary systems. Indeed, there is strong evidence that the accretion rate that produces most recycled pulsars is less than the Eddington rate (see below).

As noted above, few if any accreting neutron stars in LMXBs have spin rates higher than 750~Hz. Possible explanations for this include (see Lamb \& Yu 2005) (a)~the finite amount of angular momentum that can be accreted during binary mass transfer; (b)~the limiting effect of the magnetic braking component of the accretion torque, which becomes dominant at high spin rates; and (c)~the possible existence of one or more other braking effects---such as gravitational radiation---at high spin rates. I discuss these possibilities in turn in the following paragraphs.

(a)~Some neutron stars in LMXBs may continue to spin up throughout the mass transfer phase of binary evolution but never reach a spin rate higher than a few hundred Hz simply because the spin-up timescale given by equation~(\ref{t-s}) becomes so long that mass transfer ends before such a high spin rate can be reached (Lamb \& Yu 2005). This appears to have been overlooked in many analyses.

A few neutron stars in LMXBs are accreting steadily at rates $\sim\,$${\dot M}_E$, but most appear to be accreting at average rates $\lesssim 10^{-2}{\dot M}_E$. A star spinning at 600~Hz with an average accretion rate $\sim 10^{-3}{\dot M}_E$ has a spin-up timescale $\sim 2$~Gyr, even if the magnetic braking component of the accretion torque and emission of gravitational radiation are both negligible. This is longer than the mass transfer phase of many binary systems. In such systems, mass transfer will end before the star can be spun up further.

(b)~The spin rates of many neutron stars in LMXBs appear to be limited by the magnetic braking component of the accretion torque (Ghosh \& Lamb 1979a; Ghosh \& Lamb 1979b; Ghosh \& Lamb 1992; Lamb \& Yu 2005). The properties of accreting neutron stars in LMXBs provide some evidence for this, but the most compelling evidence for magnetic spin equilibrium is provided by the properties of the rotation-powered radio MSPs that are discussed below.

\begin {figure}[t!]
\hskip5pt\includegraphics[width=13.5cm]{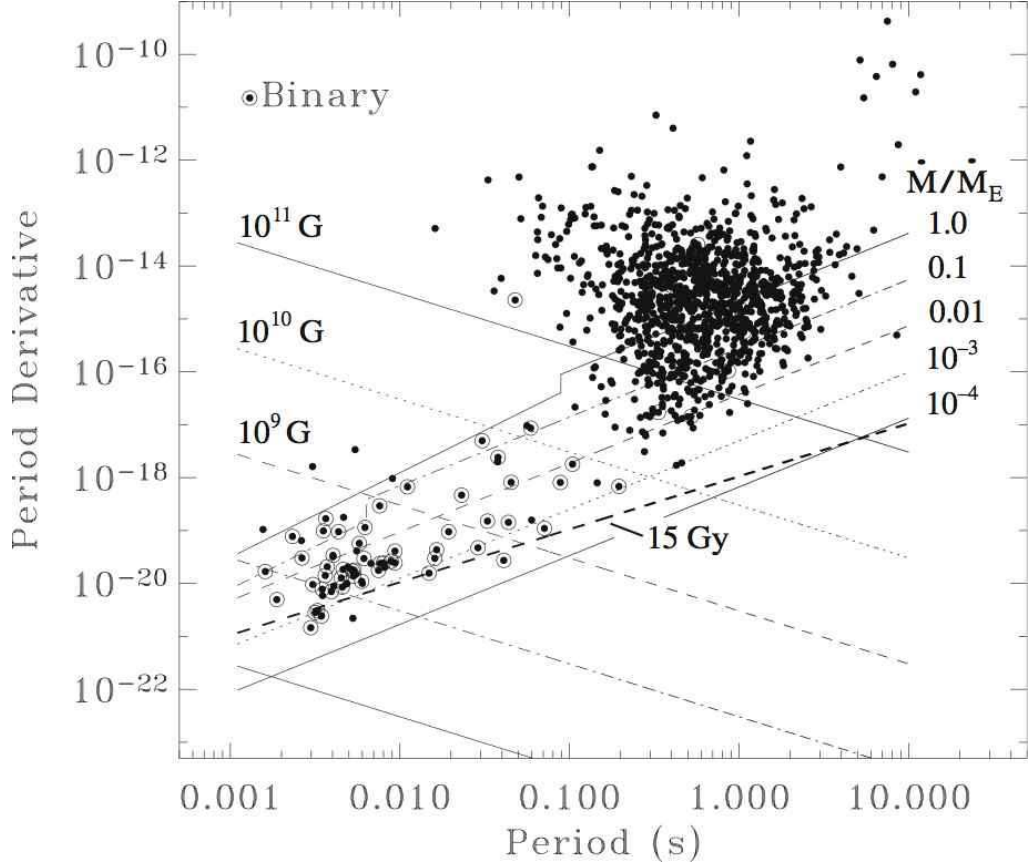}
\vskip-5pt
\caption {Pulsar evolution diagram (PED) used to analyze the evolution of accretion-powered X-ray pulsars and the production and evolution of rotation-powered radio pulsars. The data points are known rotation-powered pulsars; those of pulsars in binary systems are encircled. The data are from Hobbs \& Manchester (2004). \label{spin-ev-fig}}
\end {figure}

The spin and magnetic field evolution of accretion-powered X-ray pulsars, the production of rotation-powered radio pulsars, and the spin and magnetic field evolution of the latter can be analyzed using the pulsar-evolution diagram (PED) shown in figure~\ref{spin-ev-fig} (Ghosh \& Lamb 1992; Lamb \& Yu 2005).

The lines in the PED that slope downward to the right show the magnetic dipole braking relation $B_d = 3.2 \times 10^{19} (P_s{\dot P_s})^{1/2}$~G for a \textit {rotation-powered pulsar} with a constant magnetic field of the strength given by the labels of the lines. The heavy dashed line that slopes upward to the right shows where stars with a magnetic dipole spin-down timescale equal to 15~Gy would lie.

The lines in the PED that slope upward to the right show the equilibrium spin period $P_{\rm eq}$ specified on the horizontal axis for an \textit {accretion-powered pulsar} that is accreting at the rate given by the labels of these lines and has a dipole magnetic field that corresponds to the magnetic dipole braking rate specified on the vertical axis (the actual $\dot P_s$ of an accretion-powered pulsar that is in magnetic spin equilibrium is of course zero).

The $P_{\rm eq}$ lines in the PED for ${\dot M}={\dot M}_E$ and ${\dot M}=0.1{\dot M}_E$ have jumps where the structure of the disk changes from RPD (lower left) to GPD (upper right). In reality the transition is smooth. The lines for ${\dot M} \lesssim 0.01{\dot M}_E$ do not have jumps, because for these accretion rates the disk is GPD at $r_m$, even if the star's magnetic field is $\lesssim 3 \times 10^7$~G. The effects of the stellar surface and the innermost stable circular orbit are not shown. 

The spin rate of an accreting neutron star reflects the accretion torque acting on it, averaged over the spin relaxation time $t_{s}$ given by equation~(\ref{t-s}). As noted above, a few neutron stars in LMXBs are accreting steadily at rates $\sim {\dot M}_E$, but most are accreting at rates $\lesssim 10^{-2}{\dot M}_E$ and consequently have spin relaxation timescales $\gtrsim 10^8$ years. Determining the average accretion torque over such a long interval is difficult, because of our lack of knowledge of the accretion rate and magnetic field histories of these stars (Miller et al.\ 1998; Lamb \& Yu 2005). Many accrete only episodically. For example, the recently-discovered accretion-powered MSPs in LMXBs usually have accretion rates $\lesssim 10^{-4}{\dot M}_E$, but also have outbursts every few years. During these outbursts, which last for several weeks, their accretion rates rise to $\sim 10^{-2}{\dot M}_E$ before falling back to $\lesssim 10^{-4}{\dot M}_E$ (Chakrabarty et al.\ 2003; Strohmayer et al.\ 2003). Also, the magnetic fields of accreting neutron stars appear to decrease by factors $\sim 10^2$--$10^3$ as a result of accretion, perhaps on timescales as short as hundreds of years (see Lamb \& Yu 2005). All these variations make it difficult to determine the average accretion torque over long time intervals. If accretion torque theory can be validated, the argument can be reversed and spin-evolution measurements can be used to determine the average accretion rate over binary evolution timescales.

The spin frequencies and accretion rates of the 18 known accretion- and nuclear-powered MSPs in LMXBs are consistent with long-term magnetic spin equilibrium if their dipole fields are between $3 \times 10^7$~G and $3 \times 10^8$~G and their time-averaged accretion rates are between $3 \times 10^{-3}{\dot M}_E$ and ${\dot M}_E$. The spin rates and visible X-ray oscillations of the accretion-powered MSPs are understandable if they have magnetic fields $\sim 3 \times 10^8$~G and have been spun up to magnetic spin equilibrium by accretion at rates \mbox {$\sim 10^{-2} {\dot M}_E$}. The higher spin rates of many of the nuclear-powered MSPs are understandable if their magnetic fields are less than $\sim 10^8$~G and their time-averaged accretion rates are $\sim 10^{-2}{\dot M}_E$.

If many accreting MSPs are near magnetic spin equilibrium, the ones with lower spin rates should have higher magnetic fields and hence stronger X-ray oscillations, other things being equal, making them easier to detect. Indeed, six of the MSPs in which accretion-powered oscillations have been detected do have spin rates $<\,$500~Hz whereas only one has a spin rate $>\,$500~Hz (see Table~1), consistent with this expectation. More MSPs are needed to draw firm conclusions.

The absence of accretion-powered millisecond X-ray pulsars with frequencies $>620$~Hz is consistent with their being in time-averaged magnetic spin equilibrium, if their magnetic fields are greater than $\sim 3 \times 10^7$~G and their time-averaged accretion rates are less than \mbox{$\sim 10^{-3}{\dot M}_E$}. These fields and rates are consistent with the other observed properties of neutron stars in LMXBs (Miller et al.\ 1998; Psaltis \& Chakrabarty 1999; Chakrabarty et al.\ 2003).

(c)~The spin-up of neutron stars in LMXBs by accretion may be limited by gravitational wave emission, if they have weak enough magnetic fields and large enough time-dependent mass- or mass-current quadrupoles (Ushomirsky, Cutler \& Bildsten 2000; Lindblom \& Owen 2002). However, at present there is no unambiguous evidence for this.

We close this discussion of accretion-powered millisecond X-ray pulsars by emphasizing again that the paucity of accreting neutron stars with spin rates higher than 620~Hz may be due simply to the very long time required for accretion to produce such high spin rates (Lamb \& Yu 2005): for $\nu_{s} = 750$~Hz and a time-averaged accretion rate $\sim 10^{-3}{\dot M}_E$, the spin-up timescale is $\sim 2.5$ Gyr, longer than the mass transfer phase in low-mass systems. 

\textit {Production and spin evolution of rotation-powered millisecond radio pulsars}.---The initial spin periods of rotation-powered millisecond pulsars produced by accretion in binary systems reflect the appropriately time-averaged magnetic fields and accretion rates of the progenitor neutron stars toward the end of the mass transfer phase (Ghosh \& Lamb 1979b; Ghosh \& Lamb 1992; Lamb \& Yu 2005). Comparison of the $P_s$--${\dot P}_s$ distribution of known rotation-powered radio pulsars with the pulsar evolution diagram (PED) shown in Figure~\ref{spin-ev-fig} therefore provides important information about how they are produced:

(1)~The current positions of recycled pulsars in the PED are necessarily to the right of their positions when accretion ended, because after accretion ends they will be spun down by the magnetic braking torque, even if there is no other braking torque. If the braking torque on these pulsars is small enough, their positions in the PED will not have changed significantly since accretion ended and their current positions can be taken as indicating their positions at the end of the mass transfer phase. This is the case for pulsars near and below the 15~Gy heavy dashed line.

(2)~If the accretion-powered X-ray MSPs that are the progenitors of the rotation-powered radio MSPs are in magnetic spin-equilibrium at the time accretion ends, the standard theory of disk accretion by magnetic neutron stars predicts that there should be no rotation-powered MSPs to the left of the magnetic spin-equilibrium line for ${\dot M} = {\dot M}_E$, because (i)~this line defines the shortest spin period that can be reached by accretion and (ii)~any magnetic braking that occurs after accretion ends will drive the pulsar to the right in the PED. We emphasize that an RPD model of the inner disk must be used for neutron stars with such weak magnetic fields accreting at such high rates.

(3)~The existence of an upper edge to the distribution of rotation-powered MSPs in the PED indicates that the accretion torque vanishes at high spin rates, i.e., \textit {there is an equilibrium spin rate}. The agreement of this edge with the magnetic spin-equilibrium line predicted for ${\dot M} = {\dot M}_E$ by the standard model of disk accretion by weakly magnetic MSPs (Ghosh \& Lamb 1979b; Ghosh \& Lamb 1992) indicates that (i)~the standard model is roughly correct for stars with magnetic fields $\sim 3 \times 10^8$~G to $\sim 3 \times 10^{10}$~G accreting at ${\dot M} \approx {\dot M}_E$, (ii)~some of the progenitors of the rotation-powered MSPs had accretion rates $\approx {\dot M}_E$ for long enough to reach the magnetic spin equilibrium line for ${\dot M} = {\dot M}_E$ by the time mass transfer ended, (iii)~accretion by these stars ended abruptly enough that they did not spin down appreciably after reaching this spin equilibrium line, and (iv)~the spin rates of an appreciable fraction of accretion-powered MSPs are limited primarily by magnetic braking.

The reported $P$'s and ${\dot P}$'s of two rotation-powered MSPs recently discovered in globular clusters (\mbox {B1821$-$24} and \mbox {B1820$-$30A}; Hobbs et al.\ 2004) place them above the magnetic spin-equilibrium line predicted for ${\dot M} = {\dot M}_E$. Either the intrinsic ${\dot P}_s$'s of these pulsars are less than shown or the standard RPD model of the inner disk does not accurately describe the accretion flows that spun up these stars.

(4)~The MSPs that appear near the upper edge of the MSP distribution in the PED must have accreted at near-Eddington rates for at least $\sim1\,$Myr and therefore must have accreted at least $\sim\,$0.1$M_\odot$.

(5)~Many pulsars that are well below the magnetic spin-equilibrium line for accretion rates $\sim 10^{-2} {\dot M}_E$ have very long magnetic braking spin-down times. This implies either (i)~they never reached spin equilibrium or (ii)~they accreted at rates $\ll {\dot M}_E$ for a long time before the end of the mass transfer phase.

(6)~The magnetic spin-equilibrium hypothesis predicts that MSPs should be rare or absent below the ${\dot M} = 10^{-4} {\dot M}_E$ spin-equilibrium line in the PED, because stars accreting at such low rates generally will not achieve spin equilibrium during their accretion phase. The observed $P_s$--${\dot P}_s$ distribution is consistent with this prediction.

(7)~The MSPs near the 15~Gyr spin-down line in the PED were produced \textit {in situ} by accretion at rates $\lesssim 3 \times 10^{-3} {\dot M}_E$ rather than by spin-up to shorter periods by accretion at rates greater than $3 \times 10^{-3} {\dot M}_E$ followed by magnetic dipole braking, because braking would take too long. This result accords with the expectation noted above that most neutron stars in LMXBs accrete at rates $\ll {\dot M}_E$ toward the end of their accretion phase.

In summary, the properties of the accretion- and nuclear-powered X-ray MSPs discovered in LMXBs using \textit {RXTE} support the idea that such stars are the progenitors of the observed rotation-powered radio-emitting MSPs. After being spun down to longer periods by rotational energy loss, the X-ray MSPs discovered using \textit {RXTE} were spun up to millisecond periods by accretion, eventually becoming the nuclear- and accretion-powered MSPs we see today. Some of these may stop accreting before reaching magnetic spin equilibrium or may be spun down as accretion ends, but we expect many of them to become rotation-powered radio MSPs.

I thank L. Bildsten, D. Chakrabarty, P. Kaaret, M. van der Klis, M.~C. Miller, D. Psaltis, J. Swank, and W.~Yu for helpful discussions and W.~Yu for help in preparing the references.


\begin{thebibliography}{99}

\bibitem[Alpar et al.(1982)]{alpar-1982}
Alpar, M.~A., Cheng, A.~F., Ruderman, M.~A., \& Shaham, J., A new class of radio pulsars, Nature, 300, 728--730, 1982.
\vskip8pt

\bibitem[Backer et al.(1982)]{Backer-1982}
Backer, D.~C., Kulkarni, S.~R., Heiles, C., Davis, M.~M., \& Goss, W.~M., A millisecond pulsar, Nature, 300, 615--618, 1982.
\vskip8pt

\bibitem[Backer, Chakrabarty \& Morgan(1998)]{Backer-1998} Chakrabarty, D., \& Morgan, E. H., The two-hour orbit of a binary millisecond X-ray pulsar, Nature, 394, 346--348, 1998.
\vskip8pt

\bibitem[Chakrabarty et al.(2003)]{Chakrabarty-2003}
Chakrabarty, D., Morgan, E.~H., Muno, M.~P., Galloway, D.~K., Wijnands, R., van der Klis, M., \& Markwardt, C.~B. Nuclear-powered millisecond pulsars and the maximum spin frequency of neutron stars, Nature, 424, 42--44, 2003.
\vskip8pt

\bibitem[Galloway et al.(2002)]{galloway-2002}
Galloway, D.~K., Chakrabarty, D., Morgan, E. H., \& Remillard, R. A., Discovery of a high-latitude accreting millisecond pulsar in an ultracompact binary, ApJ, 576, L137--L140, 2002.
\vskip8pt

\bibitem[Galloway et al.(2001)]{galloway-2001}
Galloway, D.~K., Chakrabarty, D.,  Muno, M.~P., \& Savov, P., Discovery of a 270~Hertz X-ray burst oscillation in the X-ray dipper 4U~1916$-$053, ApJ, 549, L85--L88, 2001.
\vskip8pt

\bibitem[Ghosh \& Lamb(1979a)]{ghosh-lamb1979a}
Ghosh, P., \& Lamb, F.~K., Accretion by rotating magnetic neutron stars. II - Radial and vertical structure of the transition zone in disk accretion, ApJ, 232, 259--276, 1979a.
\vskip8pt

\bibitem[Ghosh \& Lamb(1979b)]{ghosh-lamb1979b}
Ghosh, P., \& Lamb, F.~K., Accretion by rotating magnetic neutron stars. III - Accretion torques and period changes in pulsating X-ray sources, ApJ, 234, 296--316, 1979b.
\vskip8pt

\bibitem[Ghosh \& Lamb(1992)]{ghosh-lamb1992}
Ghosh, P., \& Lamb, F.~K. Diagnostics of disk-magnetosphere interaction in neutron star binaries, in X-Ray Binaries and Recycled Pulsars, ed.\ E.~P.~J. van den Heuvel \& S. Rappaport (Kluwer: Dordrecht), 487--510, 1992.
\vskip8pt

\bibitem[Hartman et al.(2003)]{hartman-2003}
Hartman, J. M., Chakrabarty, D., Galloway, D. K.,  Muno, M. P., Savov, P., Mendez, M., van Straaten, S., \& Di Salvo, T., Discovery of 619~Hz thermonuclear burst oscillations in the low-mass X-ray binary 4U~1608$-$52, American Astronomical Society, HEAD Meeting No.~35, abstract~17.38, 2003.
\vskip8pt

\bibitem[Hessels et al.(2006)]{hessels-2006}
Hessels, J.~W.~T., Ransom, S.~M., Stairs, I.~H., Freire, P.~C.~C., Kaspi, V.~M., \& Camilo, F., A radio pulsar spinning at 716~Hz, Science, 311, 1901--1904, 2006.
\vskip8pt

\bibitem[Hobbs \& Manchester(2004)]{hobbs-manchester2004}
Hobbs, G.~B., \& Manchester, R.~N., ATNF Pulsar Catalogue v1.2, \break http://www.atnf.csiro.au/research/pulsar/psrcat/psrcat\_help.html, 2004.
\vskip8pt

\bibitem[Kaaret et al.(2002)]{kaaret-2002}
Kaaret, P., Zand, J.~J.~M. in't., Heise, J., \& Tomsick, J. A., Discovery of millisecond variability in the neutron star X-ray transient SAX~J1750.8$-$2900, ApJ, 575, 1018--1024, 2002.
\vskip8pt

\bibitem[Kaaret et al.(2003)]{kaaret-2003}
Kaaret, P., Zand, J.~J.~M. in't., Heise, J., \& Tomsick, J. A., Discovery of X-ray burst oscillations from a neutron star X-ray transient in the globular cluster NGC~6440, ApJ, 598, 481--485, 2003.
\vskip8pt

\bibitem[Lamb(2003)]{lamb2003}
Lamb, F.~K., High-frequency QPOs in neutron stars and black holes: Probing dense matter and strong gravitational fields, in From X-Ray Binaries to Gamma-Ray Bursts, ed.\ E.~P.~J. van den Heuvel, L.~Kaper, E.~Rol, \&  R.~A.~M.~J. Wijers (San Francisco: Astron. Soc. Pacific), 221--250, 2003.
\vskip8pt

\bibitem[Lamb(2005)]{lamb2005}
Lamb, F.~K., Millisecond X-ray pulsars and QPOs, in The Electromagnetic Spectrum of Neutron Stars, ed.\ A.~Baykal, S.~K.~Yerli, S.~C.~Inam, \& S.~Grebenev (Springer: Berlin), 311--326, 2005.
\vskip8pt

\bibitem[Lamb \& Miller(2001)]{Lamb-Miller2001}
Lamb, F.~K., \& Miller, M.~C., Changing frequency separation of kilohertz quasi-periodic oscillations in the sonic-point beat-frequency model, ApJ, 554, 1210--1215, 2001.
\vskip8pt

\bibitem[Lamb \& Miller(2006)]{lamb-miller2006}
Lamb, F.~K. \& Miller, M.~C., General relativistic sonic-point and spin-resonance beat-frequency model of the kilohertz QPO pairs, ApJ, submitted (astro-ph/0308179), 2006.
\vskip8pt

\bibitem[Lamb, Pethick \& Pines(1973)]{lamb-2003}
Lamb, F.~K., Pethick, C.~J, \& Pines, D. A model for compact X-ray sources: Accretion by rotating magnetic stars, ApJ, 184, 271--289, 1973.
\vskip8pt

\bibitem[Lamb \& Yu(2005)]{lamb-yu2005}
Lamb, F.~K., \& Yu, W., Spin rates and magnetic fields of millisecond pulsars, in Binary Radio Pulsars, ed.\ F. A. Rasio \& I. H. Stairs (ASP Conference Series, Vol. 328), 299--309, 2005.
\vskip8pt

\bibitem[Linares et al.(2005)]{linares-2005}
Linares, M., van der Klis, M., Altamirano, D., \& Markwardt, C., Discovery of kilohertz quasi-periodic oscillations and shifted frequency correlations in the accreting millisecond pulsar XTE~J1807$-$294, ApJ, 634, 1250--1260, 2005.
\vskip8pt

\bibitem[Lindblom \& Owen(2002)]{lindblom-owen2002}
Lindblom, L., \& Owen, B., Effect of hyperon bulk viscosity on neutron-star $r$-modes, Phys.\ Rev.\ D, 65, 063006, 2002.
\vskip8pt

\bibitem[Lindblom(2005)]{lindblom2005}
Lorimer, D.~R., Binary and millisecond pulsars, Living Rev.\ Relativity 8, 7. URL (cited on April 4, 2006):
http://www.livingreviews.org/lrr-2005-7, 2005.
\vskip8pt

\bibitem[Markwardt, Strohmayer \& Swank(1999)]{markwardt-1999}
Markwardt, C.~B., Strohmayer, T.~E., \& Swank, J.~H., Observation of kilohertz quasi-periodic oscillations from the atoll source 4U~1702$-$429 by the Rossi X-Ray Timing Explorer, ApJ, 512, L125--L129, 1999.
\vskip8pt

\bibitem[Markwardt, Swank \& Strohmayer(2004)]{markwardt-2004}
Markwardt, C.~B., Swank, J. H., \& Strohmayer, T. E., IGR~J00291$+$5934 is a 598~Hz X-ray Pulsar, The Astronomer's Telegram, 353, 2004.
\vskip8pt

\bibitem[Markwardt et al.(2002)]{markwardt-2002}
Markwardt, C.~B., Swank, J.~H., Strohmayer, T.~E., Zand, J.~J.~M. in't., \& Marshall, F.~E., Discovery of a second millisecond accreting pulsar: XTE~J1751$-$305, ApJ, 575, L21--L24, 2002.
\vskip8pt

\bibitem[Miller, Psaltis \& Lamb(1998)]{MPL98}
Miller, M.~C., Lamb, F.~K., \& Psaltis, D., Sonic-point model of kilohertz quasi-periodic brightness oscillations in low-mass X-ray binaries, ApJ, 508, 791--830, 1998.
\vskip8pt

\bibitem[Morgan, Kaaret \& Vanderspek(2005)]{Morgan-2005}
Morgan, E., Kaaret, P., \& Vanderspek, R., HETE~J1900.1-2455 is a millisecond pulsar, The Astronomer's Telegram, 523, 2005.
\vskip8pt

\bibitem[Psaltis \& Chakrabarty(1999)]{psaltis-chakrabarty1999}
Psaltis, D., \& Chakrabarty, D., The disk-magnetosphere interaction in the accretion-powered millisecond pulsar SAX~J1808.4-3658, ApJ, 521, 332--340, 1999.
\vskip8pt

\bibitem[Radhakrishnan \& Srinivasan(1982)]{radhakrishnan-1982}
Radhakrishnan, V., \& Srinivasan, G., On the origin of the recently discovered ultra-rapid pulsar, Curr.\ Sci., 51, 1096--1099, 1982.
\vskip8pt

\bibitem[Romanova et al.(2002)]{Romanova-2002}
Romanova, M.~M., Ustyugova, G.~V., Koldoba,  A.~V., \& Lovelace, R.~V.~E., Magnetohydrodynamic simulations of disk-magnetized star interactions in the quiescent regime: Funnel flows and angular momentum transport, ApJ, 578, 420--438, 2002.
\vskip8pt

\bibitem[Smith, Morgan \& Bradt(1997)]{smith-1997}
Smith, D.~A., Morgan, E.~H., \& Bradt, H., Rossi X-Ray Timing Explorer discovery of coherent millisecond pulsations during an X-ray burst from KS~1731$-$260, ApJ, 479, L137--L140, 1997.
\vskip8pt

\bibitem[Strohmayer et al.(1997)]{strohmayer-1997}
Strohmayer, T.~E., Jahoda, K., Giles, A.~B., \& Lee, U., Millisecond pulsations from a low-mass X-ray binary in the galactic center region, ApJ, 486, 355--362, 1997.
\vskip8pt

\bibitem[Strohmayer et al.(2003)]{Strohmayer-2003}
Strohmayer, T.~E., Markwardt, C.~M., Swank, J.~H., \& Zand, J.~J.~M. in't., X-ray bursts from the accreting millisecond pulsar XTE~J1814$-$338, ApJ, 596, L67--L70, 2003.
\vskip8pt

\bibitem[Strohmayer et al.(1996)]{Strohmayer-1996}
Strohmayer, T.~E., Zhang, W., Swank, J.~H., Smale, A.~P., Titarchuk, L., \& Day, C., Millisecond X-ray variability from an accreting neutron star system, ApJ, 469, L9--L12, 1996.
\vskip8pt

\bibitem[Strohmayer et al.(1998)]{strohmayer-1998}
Strohmayer, T.~E., Zhang, W., Swank, J.~H., White, N.~E., \& Lapidus, I., On the amplitude of burst oscillations in 4U~1636$-$54: Evidence for nuclear-powered pulsars, ApJ, 498, L135--139, 1998.
\vskip8pt

\bibitem[Taylor \& Stinebring(1986)]{taylor-stinebring1986}
Taylor, J. H. \& Stinebring, D. R., Recent progress in the understanding of pulsars, ARA\&A, 24, 285--327, 1986.
\vskip8pt

\bibitem[Ushomirsky, Cutler \& Bildsten(2000)]{ushomirsky-2000}
Ushomirsky, G., Cutler, C., \& Bildsten, L., Deformations of accreting neutron star crusts and gravitational wave emission, MNRAS, 319, 902--932, 2000.
\vskip8pt

\bibitem[van der Klis(2006)]{michiel2006}
van der Klis, M., Rapid X-ray variability, in Compact Stellar X-Ray Sources, ed. W.~H.~G. Lewin \& M.~van der Klis (Cambridge University Press: Cambridge), 39--112 (astro-ph/0410551), 2006.
\vskip8pt

\bibitem[Villareal \& Strohmayer(2004)]{Villareal-strohmayer2004}
Villarreal, A.~R., \& Strohmayer, T.~E., Discovery of the neutron star spin frequency in EXO~0748$-$676, ApJ, 614, L121--L124, 2004.
\vskip8pt

\bibitem[Wijnands, Strohmayer \& Franco(2001)]{wijnands2001}
Wijnands, R., Strohmayer, T., \& Franco, L.~M., Discovery of nearly coherent oscillations with frequency of $\sim 567$~Hz during Type-I X-ray bursts of the X-ray transient and eclipsing binary X1658$-$298, ApJ, 549, L71--L75, 2001.
\vskip8pt

\bibitem[Wijnands \& van der Klis(1998)]{rudy-michiel1998}
Wijnands, R., \& van der Klis, M., A millisecond pulsar in an X-ray binary system, Nature, 394, 344--346, 1998.
\vskip8pt

\bibitem[Wijnands et al.(2003)]{wijnands-2003}
Wijnands, R., van der Klis, M., Homan, J., Chakrabarty, D., Markwardt, C.~B., \& Morgan, E.~H., Quasi-periodic X-ray brightness fluctuations in an accreting millisecond pulsar, Nature, 424, 44--47, 2003.
\vskip8pt

\bibitem[Zhang et al.(1998)]{zhang-1998a}
Zhang, W., Jahoda, K., Kelley, R.~L., Strohmayer, T.~E., Swank, J.~H., \& Zhang, S.~N., Millisecond oscillations in the persistent and bursting flux of Aquila~X-1 during an outburst, ApJ, 495, L9--L12, 1998.
\vskip8pt

\bibitem[Zhang et al.(1998)]{}
Zhang, W., Lapidus, I., Swank, J.~H., White, N.~E., \&  Titarchuk, L., 4U~1636$-$53x, IAU Circ.~6541, 1998.

\end{thebibliography}
\end{document}